\documentclass{aa}

\usepackage{graphicx}
\usepackage{txfonts}
\usepackage{natbib}
\usepackage{ulem}
\begin{document}

\title{Spectropolarimetric multi line analysis of stellar magnetic fields}

\author{ {J.C. Ram\'{\i}rez V\'elez } \inst{1,2},
 { M. Semel }\inst{2},  M. Stift  \inst{3,4}, M.J. Mart\'{\i}nez Gonz\'alez \inst{4,5}, P. Petit \inst{6} and N. Dunstone\inst{7}}
\institute{Instituto de Astronomia - UNAM, 04510, Coyoacan, DF, Mexico. \email{jramirez@astroscu.unam.mx}
\and  LESIA, Observatoire de Paris Meudon. 92195 Meudon, France. \email{Julio.Ramirez@obspm.fr, Meir.Semel@obspm.fr}
\and Institute for Astronomy, Univ. of Vienna,  T\"urkenschanzstrazze 17, 1180. Wien, Austria  \email{stift@astro.univie.ac.at}
\and LERMA, Observatoire de Paris Meudon. 92195 Meudon, France. \email{Marian.Martinez@obspm.fr}
\and Instituto de Astrofisica de Canarias, via lactea s/n, 38205, Tenerfie, Spain \email{marian@iac.es}
\and {Laboratoire d'Astrophysique Toulouse-Tarbes, U. de Toulouse, CNRS, France. \email{petit@ast.obs-mip.fr}}
 \and {School of Physics and Astronomy, U. of St Andrews, KY169SS, UK. \email{njd2@st-andrews.ac.uk} }}

\offprints{jramirez@astroscu.unam.mx}
\date{}

\begin{abstract}
{}
{In this paper we study the feasibility of inferring the magnetic field from polarized multi line spectra using two methods: \textit{  The pseudo line approach} and \textit{  The PCA-ZDI approach}.}
{We use multi-line techniques, meaning that all the lines of a stellar spectrum  
 contribute to obtain  a polarization signature. The use of multiple lines dramatically increases the signal to noise ratio of these polarizations signatures.  Using one technique,  \textit{ the pseudo line approach}, we construct the pseudo line as the mean  profile of all the individual lines. 
The other technique,  \textit{  the PCA-ZDI approach} proposed recently by Semel et al. (2006) for  the detection of  polarized signals, combines Principle Components Analysis (PCA) and the Zeeman Doppler Imaging technique (ZDI). This new method has a main advantage: the polarized signature is extracted using cross correlations between the stellar spectra and functions containing the polarization properties of each line.
These functions are the principal components of a database of synthetic spectra. The  synthesis of the spectra of the database are obtained using the radiative transfer equations in LTE. The profiles built with the PCA-ZDI technique are denominated Multi-Zeeman-Signatures.}
{The construction of the pseudo line as well as the Multi-Zeeman-Signatures  is a powerful tool in the study  of stellar and solar magnetic fields. The information of the physical parameters that governs the line formation is contained in the final polarized profiles.  In particular, using inversion codes, we have shown  that the magnetic field vector can be properly inferred with both  approaches despite the magnetic field regime.}

\end{abstract}


\titlerunning{Study of magnetic fields from multi line analysis}
\authorrunning{Ram\' irez V\'elez  et al.\ \ }
\maketitle

\section{Introduction}

The majority of cool stars, including our Sun,  have magnetically confined atmospheres. The study of the  magnetic activity is important for understanding: 1) the  dynamo effect, responsible for generating solar and stellar magnetic fields, and 2) the different states of stellar evolution and the influence of the magnetic field during these stages of evolution.

The development of the Zeeman Doppler Imaging technique, described in a series of five papers, is a benchmark   
in the stellar spectropolarimetry domain and, consequently, in the study of magnetic stars by observational methods.  It opened a new window in the research  of magnetic activity in cool  stars through the measurement of the  circular polarization  directly from the spectra. The circular and also the linear polarized signals are detectable through a combination of the Doppler and Zeeman effects \citep{zdi_1}.

Few years after the development of the ZDI technique, Semel \& Li (1996) proposed the line addition principle to improve the signal-to-noise ratio and make possible the detection  of extremely faint polarization signals in stars. Since then, the addition principle has been used in  different and more sophisticated techniques, and it  has been particularly successful with the  Least Squares Deconvolution (LSD) technique  \citep{lsd},  to construct mean  intensity  and/or polarization profiles. These  mean profiles   serves in a secondary step as input for codes that  produce magnetic stellar surface maps. Observations at different phases of star's rotational period are required  to  produce these maps \citep{zdi_4, zdi_5,  zdi_codes}.

The simplest way to employ the line addition principle is to average the spectral lines.  However, this brut line addition technique is not currently used to retrieve the magnetic field vector, not even for solar observations,  where the inversion of the Stokes parameters in individual spectral lines is a commonly employed technique. Semel et al. (2009) have recently shown that the signal obtained from the average of multiple spectral lines, the so-called pseudo line,  can be employed to estimate the longitudinal  magnetic field using  the center of gravity method (e.g. Rees et al. 1979).

In this work we present, in Sect. 2,  more accurate inversion methods applied to the multi line approach by means of the construction and inversion of the  pseudo line, and we extend the study to the linear states of polarization since considering the Stokes parameters (Q,U)  is necessary to fully determine the  orientation of the magnetic field. 

The second part of this work, Sect. 3, is dedicated to another approach of detection of magnetic fields in stars, where  all the lines contained in a given spectral range contribute to the final Stokes profiles. We first present the basis of the so-called PCA-ZDI technique. We then  describe the procedure to obtain the final Stokes profiles named Multi-Zeeman-Signatures. We then apply the developed technique to observed spectra of three cool stars.  Finally,  using synthetic spectra, we will show that  the  stellar magnetic fields can be correctly retrieved performing direct inversions of the  Multi-Zeeman-Signature profiles. 

The general conclusions are presented in Sect. 4.


\section{The pseudo line approach}

In this Sect.  we concentrate on the line addition technique using the simplest way of combine multiple spectral lines that is averaging them. We have named the  \emph{pseudo-line} ($PL$)  to the resulting  mean signal because it  does not come from any physical entity, i.e. it has not associated any intrinsic atomic property like Lande factor, potentials of excitation, etcetera. 

The Stokes vector of the pseudo line  , $\vec{S}_{PL}=(I,Q,U,V)$,  is  obtained averaging by separate in each  Stokes parameter, 

\begin{equation}
\vec{S}_{PL}=\ \sum_{i=1}^{n_l} \frac{ \left[  I_i(\lambda_R), \ \ 
 Q_i(\lambda_R), \ \  U_i(\lambda_R), \ \  V_i(\lambda_R) \right] } {n_l}, 
\label{equ:ps}
\end{equation}
where  $\lambda_R$ = $\lambda$ - $\lambda_c$ denotes the reduced wavelength and ${n_l}$ is the number of spectral lines.

 In table \ref{tab:tab1} are listed the individual  lines that we have included  in this work for the construction of the pseudo line.  They have be chosen to be the same lines as in our previous work  Semel et al. (2009), hereafter referred as Paper I.

Considering the solar case of a single point on the stellar surface,  we used the code {\sc diagonal} (L\'opez Ariste \& Semel 1999) to compute the Stokes profiles of the lines listed in table \ref{tab:tab1}, and consequently,  to obtain  the profiles of the pseudo line. 

As mentioned in Paper I, we will assume that there is only one magnetic field vector on the stellar surface  responsible of the polarized signals and that the total radiation comes from this magnetic. We do not ignore that more sophisticated scenarios can be considered, but we are not interested in simulate a real stellar spectra but in show that the capabilities of the pseudo line to retrieve the magnetic field vector.

\begin{figure*}[htpb]
\begin{center}
\resizebox{18cm}{!}{\includegraphics{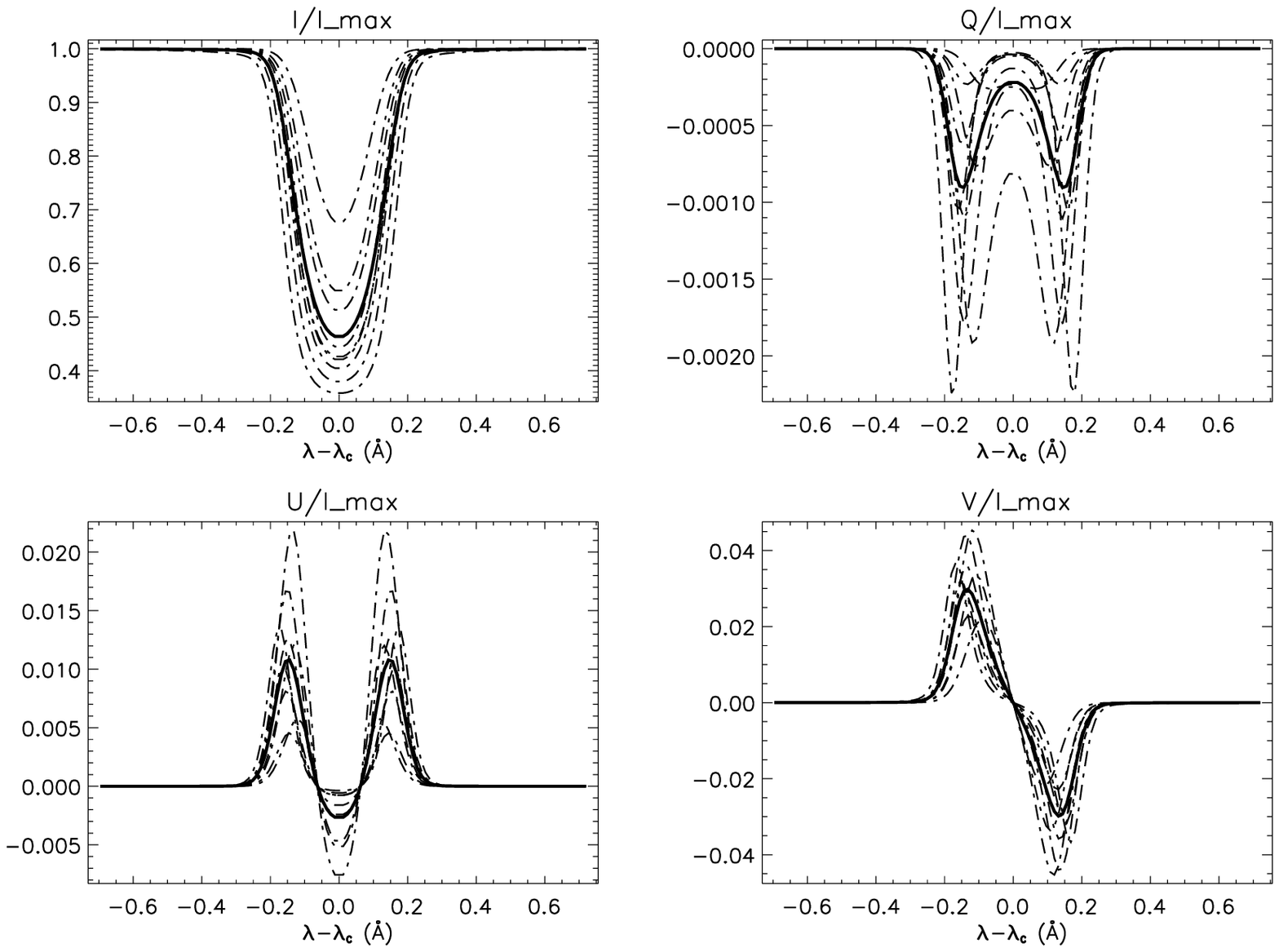}{\includegraphics{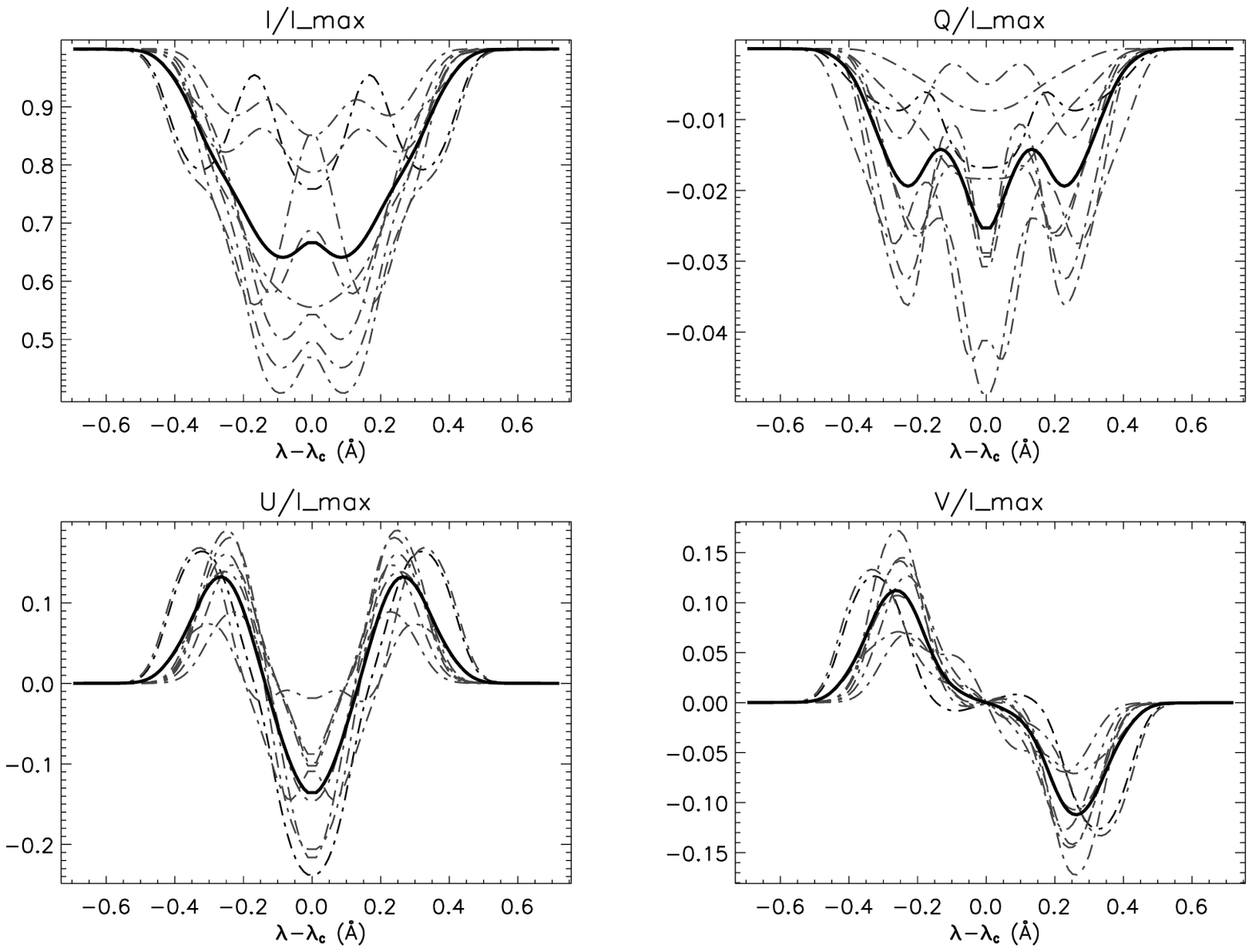}}}
\caption{The dashed lines represent  the Stokes profiles of each of the individual lines listed in table 1. The solid line represents the pseudo line obtained from the rest of the lines. 
\emph{Left side:} Stokes profiles considering a moderate field strength of 400 G.  \emph{Right side:} The same as in the left panel but considering a strong  field strength of 4 kG. In both cases the geometry of the magnetic field is  azimuthal and inclination angles of 70 and 45 degrees respectively.}
\label{fig:b_strong}
\end{center}
\end{figure*}

In Fig. \ref{fig:b_strong}, we show two examples of the pseudo profiles computed using the same given atmospheric model (see Sect. 2.3),  but using different magnetic strength field. In the left panel, considering a field strength of 400G, the profiles of all the individual lines are shape similarly,  and consequently,  the pseudo profiles keep the same shape. In the right panels, considering  a field strength of 4kG, the similarity in the shape of the individuals profiles is not preserved. Note for instance the variety of shape in the I and Q profiles.

The relatively reduced number of individual lines used to construct the pseudo line is a first step for a future generalization where much more lines can be included. 

Before to continue with the inspection of the pseudo line we present the procedure that we follow to express the spectral lines in function of the Doppler coordinates. 

\begin{table}[b]
\caption{Central wavelength, transition levels, relative intensities and Lande factors of the spectral lines of the multiplet 816 of the neutral iron. The relative intensities are taken from Allen (2000). }
\label{tab:tab1}
\centering
\begin{tabular}{ccccccc }
\hline
Line & $\lambda_c$ & Upper & Lower   &  Relative  & Lande \\ 
Number  & (\AA) & level & level & intensity & factor \\
\hline 
1 & 6141.73 & 5D2 & 5P3 & 1    & 1.81 \\
2 & 6232.66 & 5D1 & 5P2 & 2.25 & 1.99 \\
3 & 6246.33 & 5D3 & 5P3 & 7    & 1.58 \\
4 & 6301.51 & 5D3 & 5P2 & 8.75 & 1.66 \\
5 & 6302.50 & 5D0 & 5P1 & 3    & 2.48 \\
6 & 6336.83 & 5D1 & 5P1 & 6.75 & 2.00 \\
7 & 6400.01 & 5D4 & 5P3 & 27   & 1.27 \\
8 & 6408.03 & 5D2 & 5P1 & 5.25 & 1.01 \\
9 & 6411.65 & 5D3 & 5P2 & 14   & 1.18 \\
\hline
\end{tabular}
\end{table}

\subsection{Variable transformation and adequate spectral resolution}

In this Sect. we describe a variable transformation from the  wavelength coordinate into the velocity one: $\vec S(\lambda) \rightarrow \vec S(X)$. 

At present, the typical spectrographes used for  stellar spectropolarimetry are cross-dispersed type and cover  a range of thousands of \AA.   We will employ in our analysis a discrete selection of lines,  but  we present the general procedure as if we were dealing with  the complete spectral range.

Consider the transformation equation

\begin{equation}
X=c \ log(\lambda / \lambda_0)  
\label{ecu:cv1}
\end{equation} 
where X is expressed in $km/s$, and $\lambda$ in \AA.  $c$ is  the speed of light  and $\lambda_0$ the minimum  wavelength value in a given spectral interval. Deriving the precedent Eq., 

\begin{equation} 
dX = c  \frac{d\lambda }{\lambda},   
\end{equation} 
is possible to obtain a constant step in $km/s$ along the spectra when the  velocity values are discretized: $dX \rightarrow \Delta X$.

We have fixed the  constant step in $\Delta X$  to 
1$km/s$, such that we can invert  the equation  (\ref{ecu:cv1}) to  
express $\lambda$ as function of X as:

\begin{equation}  
\lambda=\lambda_0 exp(X/c)
\label{ecu:cv2}
\end{equation}
where X=[0,1,2,...].

Let $\lambda_{obs}$ denote the discrete wavelength values   obtained in the spectra after the data reduction.  By imposing a constant step in $\Delta X$ one  problem arises  for certain values because the relation $\lambda (X)$  given by  Eq. (\ref{ecu:cv2}) will not necessarily coincide with the values of $\lambda_{obs}$. An interpolation is then applied to determine the exact value at the wavelength of inte\-rest. Let $\lambda_n$ 
be one of the inexact values such that $\lambda_{obs,i}$ $<$ $\lambda_n$ $<$ $\lambda_{obs,i+1}$,  and let $S(X_n)$ denote any of the Stokes parameters at wavelength $\lambda_n$. The  value  $S(X_n)$ will then be calculated by:

 \begin{equation}
S(X_n) =  S(\lambda_i) + \frac{S(\lambda_{i+1})-S(\lambda_i)}
{\lambda_{i+1}- \lambda_i } (\lambda_n-\lambda_i )
\label{equ:intrapola}
\end{equation}

With this algorithm, the spectra $S(\lambda)$,  observed or synthetic, will be transformed to as function of the new variable   $S(X)$.

\subsection{The atmospheric model}

In what follows of this Sect. we explore the use of accurate inversions methods to apply to the pseudo line, i.e.  inversions techniques   based on  the radiative transfer equation (RTE).  

From experience in the solar data analysis, we know that some correlation between the atmospheric parameters  can generate ambiguities in the determination of the magnetic field when the noise level is close to the polarized signal levels, (e.g. del Toro Iniesta \& Ruiz Cobos 1996; Bellot Rubio \& Collados 2003; Mart\' inez Gonz\'alez et al. 2006,  and references therein). For simplicity, we initially avoid  any possible model parametric degeneration by assuming that the atmospheric model is known and only the magnetic field vector is unknown. With this premise, we test the inference  of the magnetic field through the inversion of the pseudo profiles under an ideal scenario where  all the variables in the RTE are related to the magnetic field vector :  The  field strength (B), the azimuthal angle ($\gamma_{azi}$) and the inclination field  ($\gamma_{incl}$). The atmospheric model we chose to compute the Stokes profiles is as follows: we fixed the gradient of the source function ($\Delta \tau = 10$), the Doppler  width  to 50 m\AA,  the ratio of the line to continuum absorption ($\eta$) is twice the relative strength of the components of the iron multiplet  and the damping parameter value is fixed to  0.01.

The four Stokes parameters for all the individual Fe lines are computed considering the same atmospheric model and the values of the magnetic field varies randomly in the following ranges : B=[0,10] kG, $\gamma_{azi}$=[0,180] degrees and $\gamma_{incl}=[0,90]$ degrees. 

Consider now a given combination of the magnetic field parameters,  $Y^n$=($B^n$,  $\gamma_{azi}^{\  n}$, $\gamma_{incl}^{\  n}$). From the synthesis of the Fe lines  for  this  combination of parameters, we then obtain the respective pseudo line for this $Y^n$ model. Note that we are in this way  \emph{associating} a magnetic model to the pseudo line. 

\begin{figure*}[!hbt]
\begin{center}
\resizebox{17cm}{!}{\includegraphics{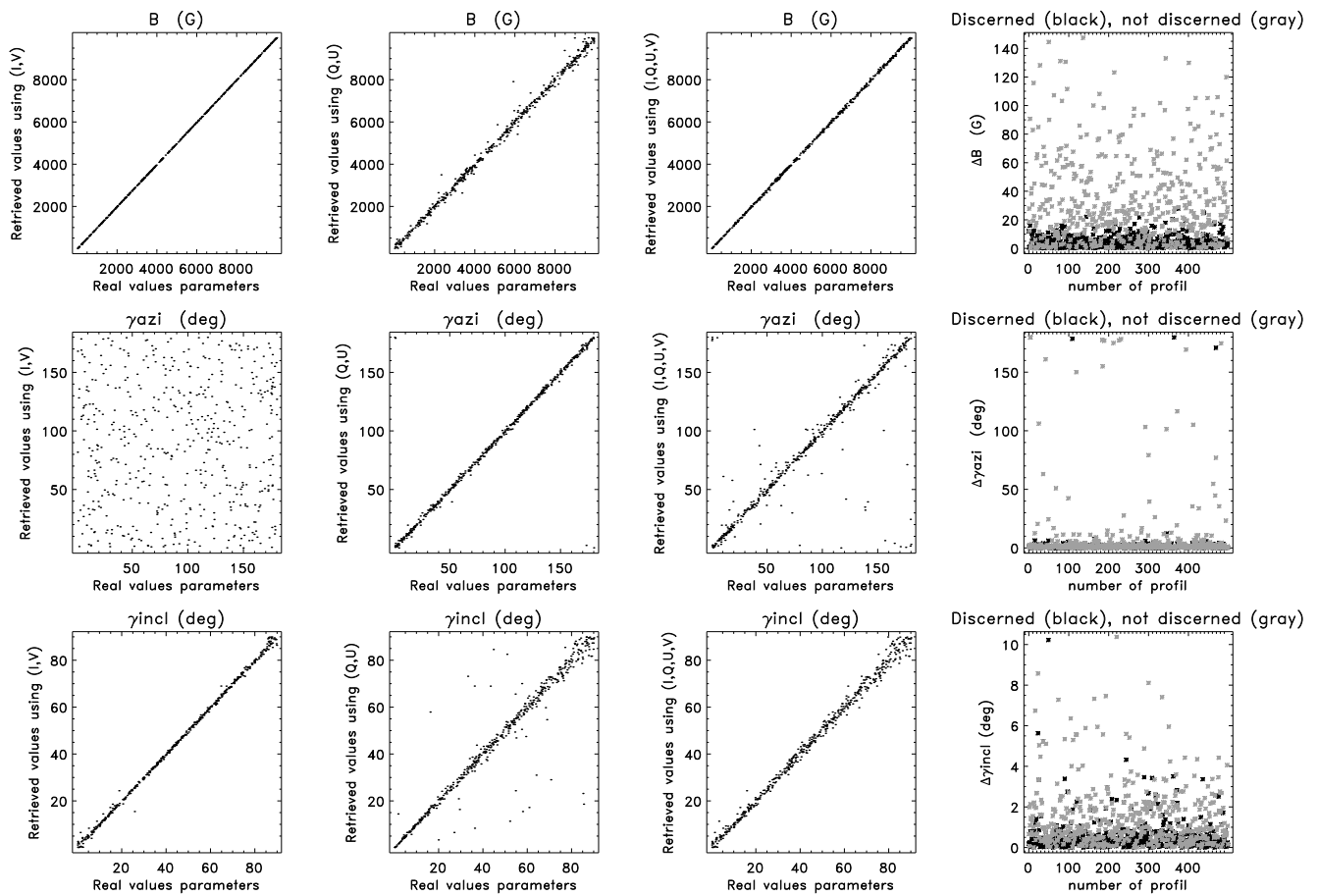}}
\caption{First and second columns: Inversions performed using different Stokes parameters (indicated in the Y-axis).  Due  to the inclusion of the magneto-optical effects in the computation of the profiles, the solutions in the second column contain the correct information in $\gamma_{incl}$  parameters but with wider ranges of errors than in the  first column. Third column: All the Stokes parameters are used to retrieve the magnetic field values. Last column: Comparison of the error inversions using the discerning criteria (in black)  and with out it (in gray), more details are given in the text. 
The better results correspond to the discerned inversions.}
\label{fig:disc}
\end{center}
\end{figure*}


\subsection{The inversion code}

The exercises consist to determine the magnetic field supposed punctual and fixed in a known position in the stellar surface. The inversions in each Fe line and in the pseudo line have been done separately and the  goal is to compare the results.

We have developed  one inversion code per Fe line and one for the pseudo line. The code's functionality is : First, we construct a representative synthetic database of a large  number of profiles (56600). Then, given a  Stokes vector to invert, we find the  best-fit solution from the database using the same algorithm as described in Ram\' irez V\'elez et al. (2008).

We inverted a set of 500 profiles for each  of the lines listed in Table  \ref{tab:tab1}.  To compute the sets of  profiles to invert, we followed the same procedure  as when constructing the database profiles:  we have fixed  the atmospheric model  and we consider random magnetic field vectors.

While inverting the set of Stokes profiles, we applied and compared different strategies to recover the magnetic field vector. Initially, we considered  the four Stokes parameters (I,Q,U,V) to simultaneously retrieve  the solution of the three magnetic parameters ($B$,  $\gamma_{azi}$,  $\gamma_{incl}$). We then inverted  the same set of profiles using only the I and V  Stokes profiles, which is the most common case for stellar observations. Finally,   we performed the inversions using only the two linear Stokes parameters (Q and U).

The goal of performing  different inversion strategies is twofold. On one hand, we compare the degree of accuracy of the inversions using the four Stokes parameters to the most typical case in the data analysis that is when we only dispose of the intensity and circular polarization spectra.

On the other hand, this procedure permits to improve the precision of the inversions.  Let us call  the \emph{discerning criteria} as the use of different Stokes parameters to retrieve the different components of the magnetic field. Given that the inversion code works finding as solution the  best-fit of the profiles in the databases, the employment of the  \emph{discerning criteria} has the advantage that the databases become largers and thus, the code reduce the error of the inversions. 
 
We remark however that the results of the analysis of the pseudo line do not depend in the proposed criteria described below, such that some readers may want to skip the next subsection.
 
\subsubsection{Database extension}

To clarify the idea of the extension of the database,  let $\vec{P_n}$ denote the Stokes vector for a given set of parameters  such that   $\vec{P_n} = \vec{P_n} (Y^n) $ and $Y^n$=($B^n$,  $\gamma_{azi}^{\  n}$, $\gamma_{incl}^{\  n}$).

Let  ($B^j$, $\gamma_{incl}^{\ j}$) denote the associated solution values found with the inversion performed using the  I and V Stokes parameters. Now, let   $\gamma_{azi}^{\  k}$ be the  solution value after  inversion of the Q and U Stokes parameters of the same  $\vec{P_n}$.  Since the values of the parameters were retrieved independently, then $j \ne k$ (in more than 90\% of the cases). Finally, we use the  so-called \emph{discerned criteria} to construct the final solution  with a parameters combination   $Y^{sol}$=($B^j$,  $\gamma_{azi}^{\  k}$, $\gamma_{incl}^{\  j}$).  Since $Y^{sol}$ was not originally in the database, we  are ``extending'' the  database through the use of the discerning criteria.

In Fig. \ref{fig:disc}, using a set of 500 profiles without any added noise, we show the inversion results following the described  criteria. The graphics correspond to the inversions of the  Fe I line at 6141 \AA, the  first line listed in Table \ref{tab:tab1}.

While using  only the I and V Stokes profiles to perform the inversions, first column in Fig. \ref{fig:disc},  more refined inversions are obtained for  $B$ and  $\gamma_{incl}$  that when all the Stokes parameters are included in the inversions (third column). This is explained because the number of magnetic models whom  parameters $B$ and  $\gamma_{incl}$  are close to the original pair of value parameters ($B^{ori}$, $\gamma_{incl}^{ori}$) get increased when the $\gamma_{azi}$ is ignored. 

In the second column we present the results of the inversions  using only the linear Stokes profiles (Q,U). In this case,  also an improvement  happens for the inversions of the parameter $\gamma_{azi}$ compared to the inversions of the third column when the four Stokes parameters are considered. 
 
From the comparison of the inversions, showed in the forth column, we then conclude that the discerning criteria improves the inversion results: In the case of the field strength,  the dispersions in the errors  diminished considerable  and the  maximum error value decreased  from $\sim$140G (symbols in gray) to $\sim $25G (symbols in black). In the case of  the  geometrical parameters, the angles  $\gamma_{incl}$ and $\gamma_{azi}$,  the inversions also improve with the adopted criteria. 

\begin{figure*}[!ht]
\begin{center}
\resizebox{14cm}{!}{\includegraphics{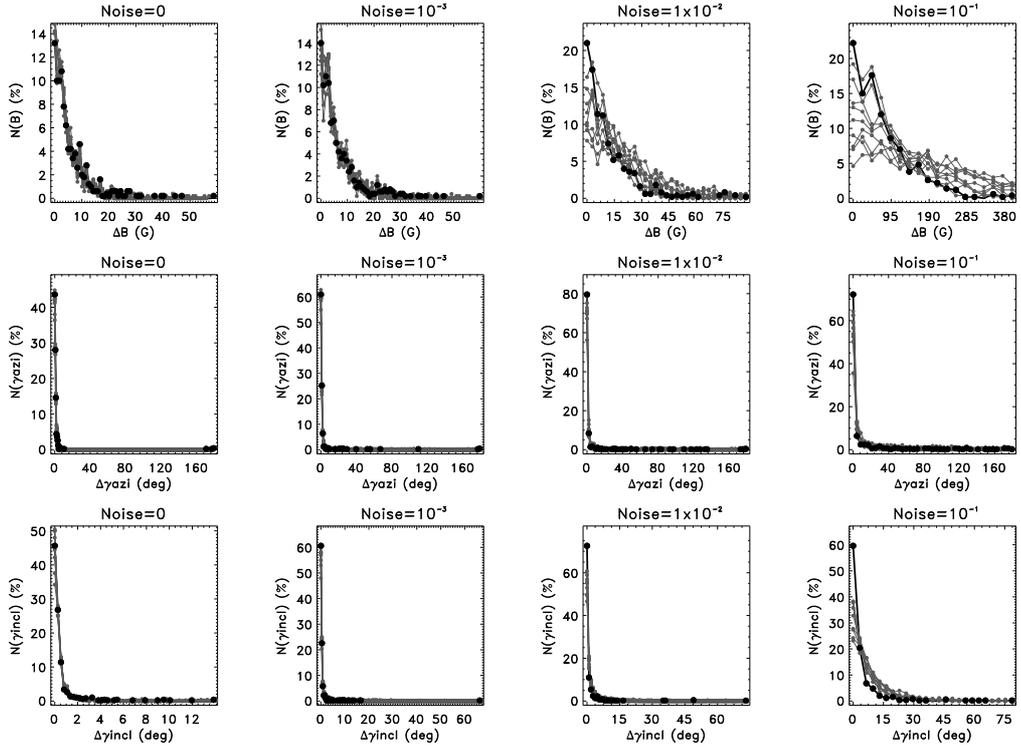}}
\caption{Percentage histograms inversions of the Fe lines (in gray color) 
and of the pseudo line (in black color).  N represents the percentage 
number of inversion cases.
\emph {In the row direction,} ordered by parameter of the magnetic 
field configuration: the strength, the azimuthal and, inclination angles.
\emph {In the column direction,}
the comparison of the inversion by noise level.} 
\label{fig:histos}
\end{center}
\end{figure*}

Note that it is expected that the proposed discerning criteria  increase the inversion precision  only for those codes that employs a big set of profiles (databases) where the solution is found.

\subsection{Testing the inversions of the pseudo line}

One advantage of the developed inversion  codes is the estimation of the typical errors. Thus, in order to compare quantitatively the inversions retrieved from each line, the results are presented in terms  of the absolute value of the difference between the original and the inferred  parameters: 

\begin{equation}
\label{deltab}
\vec{B}^{ori} -  \vec{B}^{sol}=(| \Delta B |, \ | \Delta \gamma_{azi} |, \ 
| \Delta \gamma_{incl}|).
\end{equation}

Additionally, in order to study the effect of the noise in the determination  of the magnetic field, we add to the profiles three different noise levels with respective values of  1x10$^{-3}$, 1x10$^{-2}$ and  1x10$^{-1}$ of the continuum level ($I_c$). 

In Fig.  \ref{fig:histos}  we show  the percentage histograms of the inversion's errors for all the Fe lines  (in gray) and  for  the pseudo  line (in black).  

In the case without any added noise, we can see that all the inverted parameters are well retrieved. The histograms of the errors have a sharp peak at 0 and drop quickly. Note for instance that the  field  strength is inferred with an error smaller than 10 G with a high probability. The small errors for all the parameters are mainly due to the finiteness of the database, and are identified as the precision errors of the database. In any case, note that  the results with the pseudo line are the very same  the  ones obtained from the inversions of individual spectral lines. We then conclude that the pseudo line encodes the information of the magnetic field vector as the individual spectral lines do!.

In the lowest noise case ($10^{-3} I_c$), all parameters are also correctly retrieved, the errors being very close to the precision of the database. However, when increasing the noise levels to  $10^{-2}$ and to $10^{-1} I_c$,  the error bars become more important, making the error  distributions slightly wider.  We remark that for these noisy cases the pseudo line is the one that gives the best results, corroborating  that the addition of multiple lines increases the signal-to-noise (S/N) ratio. 

\subsection{Inversions with an unknown atmospheric model}

We continue our inspection of the pseudo line relaxing  the  constraint that the magnetic field is the only free variable.  We have thus included some line formation parameters in the atmospheric model,  to verify that  in a more general case   the goodness of the approach remains.

Considering as free variables the Doppler width ($V_D$), the absorption line ratio ($\eta_0$), the source function gradient   ($\nabla \tau$)  and, the magnetic field vector, we have repeated the same exercise as before inverting a set of 500 pseudo profiles. 

The construction of the pseudo line follows the same principle as before: Given  a combination of the atmospheric parameters (excepting  $\eta_0$) those values  are used to compute  all the Fe lines, and subsequently,  obtain the pseudo line.  In the case of the  $\eta_0$ parameter, a factor  (denoted $\eta_f$) modifies in the same amount any of the $\eta_0$ values, preserving the relative intensities (i.e. for all  Fe lines,  $\eta_0$ = $\eta_0 \cdot  \eta_f$). 

In Fig.  4, we show the  obtained  results. After inversion of the 500 pseudo profiles,  we have found that  the retrieved values in any of the  parameter are correct (with of course and associated error).  We then conclude  that the  atmospheric model and the magnetic field can  both be deduced from the inversions of the pseudo line. We argue in favour of the results, that  the method used to invert the pseudo line is the very same the one used for the construction.

With the present results, we then conclude that the multi line approach based in the direct addition of many spectral lines, i.e. in the construction of the pseudo line,  is a very useful tool in the study of solar and stellar magnetic fields.

\begin{figure*}[!ht]
\begin{center}
\resizebox{16cm}{!}{\includegraphics{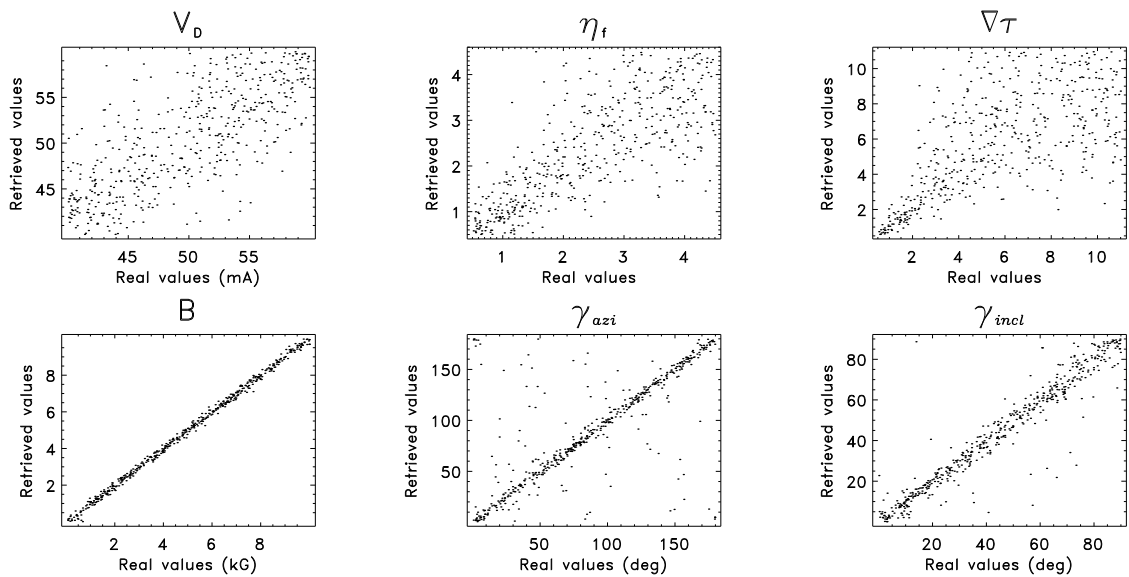}}
\vspace{-3.5cm}
\caption{Inversions of a set of 500 pseudo lines performed using the four Stokes parameters.  The title in each graphic indicates the atmospheric parameters.} 
\label{fig:graf_discern_fe}
\end{center}
\end{figure*}

\section{The PCA-ZDI approach}

Alternatively to the method presented in the last Sect., instead of make more efforts  in this direction considering as many lines as possible in the pseudo line,  we prefer  to employ a  more sophisticated  approach  recently proposed by Semel et al. (2006), hereafter referred as Paper II. In this Sect. we focus in this second approach named PCA-ZDI.

The principal idea of the PCA-ZDI technique  is to incorporate the computation of the radiative transfer equations, for instance using an atmospheric model in Local Thermodynamic Equilibrium (LTE), in combination with a powerful statistical tool, the principal components analysis (PCA), in order to  retrieve reliable detections of the stellar magnetic fields. This is in fact a way to overcome the employment of the weak field approximation used  originally in the formulation of the ZDI technique (Semel 1989) and  preserved in the LSD  technique \cite{lsd}.

\subsection{Detection of stellar  magnetic fields with PCA-ZDI}

As will be shown, the PCA-ZDI technique is a powerful method for detection of magnetic fields in stars.  In a summarized description of the developed technique, firstly is required to dispose of a representative database composed of stellar spectra. In fact, despite the domain of application, the construction of the database is a key step when is intended to perform analysis by principal components  \cite{rees_pca_00, socas_01}.  In any case, once defined the database, of what ever number of spectra is constructed  this one, it is decomposed in a new basis  using the single value algorithm \cite{golub_96}. The new  basis is spanned in terms of eigenvectors which present many interesting properties. For the purposes of this work, we employ the eigenvectors of the new  basis as \textit{detectors} to retrieve from the spectra the magnetic \textit{signature} of the Stokes parameters. We have denominated these magnetic signatures as \textit{Multi-Zeeman-Signature} (MZS). The PCA-ZDI technique will thus retrieve a not null MZS profile in the polarized Stokes parameters whenever is present a magnetic field in the stellar atmosphere and the signal to noise ratio allows it. 

From an observational point of view, the main constrain to retrieve the polarized Stokes parameters  in individual spectral lines is the very faint levels expected in the signals 

\subsection{Magnetic detectors}

 Based in  the ideas proposed in the first paper of ZDI \cite{zdi_1}, we have considered  that there is only one magnetic element in the stellar surface, at the projected  position ($\mu$), and we have calculated  the local Stokes profiles at this position. From the results already obtained with synthetic and observed spectra, we advance that the employment of the local profiles does not limit the detection in  stars with complex  magnetic fields  configurations -dipolar, multi-polar  and/or multispot-. A more detailed inspection of this statement will be presented in a forthcoming paper.   

The stellar spectra in the four Stokes parameters  have been established with the help of {\sc cossam}\footnote{Codice per la Sintesi Spettrale nelle Atmosphere Magnetiche}, described in Stift (2000) and in Wade et al. (2001). {\sc cossam} is an LTE line synthesis code in polarised light that calculates the Stokes  profiles over arbitrarily large wavelength intervals, both for the sun and for dipolar field geometries in magnetic  stars. Direct opacity sampling is carried out over the $\sigma_{-}$, the $\sigma_{+}$, and the $\pi $ components separately of the (anomalous) Zeeman patterns of the individual lines.

While covering different positions in the stellar surface and at each one considering different  field strengths and orientations is possible to obtain a representative combination of synthetic spectra which will serve to construct the database. 

The position of the stellar element  is specified through the    $\mu$=cos($\theta_{LOS}$), where $\theta_{LOS}$ is the angle  between the line-of-sight direction and the normal to surface element. $\mu$ varies from  [0.2, 1] with  a step size of 0.2. At each  $\mu$ position, we have spanned the magnetic field strength B=[0,250,...,3000] G and the inclination angle $\gamma_{incl}$=[0,15,...,90] $^{\circ}$. For economy in CPU-time we have fixed the azimuthal angle ($\gamma_{azi}$), but note that this fact does not limit the detection in the linear Stokes parameters since applying rotations of the reference frame to the calculated spectra is possible to pass from the  Stokes Q to the Stokes U,  or to obtain any desired  configuration in the linear  parameters, (e.g. Sect. 6.4 in del Toro Iniesta 2003). Consequently is justified to consider as the general case the approach  used in the $\gamma_{azi}$ parameter. In this way, the combination of parameters produces  a database ($\mathcal M$) with 425 spectra, per Stokes parameter.


The  spectral range covered from 450 to 750 nm, in steps of  10 m\AA, producing  spectra of 300,000 wavelength points in each of the Stokes parameter. In the following, we consider the Stokes spectra as  vectors $\vec{S}$ of dimension [300000].

Finally, three different atmospheric models have been considered T=3500, 4750 and  5750 $^{\circ}$K. For each model we construct the respective database. In other words, $T$  is fixed to the value $T_0$ in each database.  In all cases, we have considered the same atomic solar abundances as in Grevesse \& Sauval (1998), leaving out  the molecular solar lines.

Let $\xi_m$ represent a given combination of the  model parameters

\begin{equation}
\xi_m = \xi_m(B,\gamma_{incl}, \gamma_{azi}, \mu, T_0)_m, \ \ \ m=[0,1,..,424]
\label{equ:param}
\end{equation}
and let $\vec{S}_m(\xi_m$)  denotes the $mth$ combination in any of the Stokes parameters  calculated with {\sc cossam}. Grouping  all the 425 considered spectra models, we obtain an array  $\mathcal M$ of dimensions [300000, 425] per Stokes parameter. 

Applying the {\it Single Value Decomposition}  (SVD) procedure \cite{golub_96},  the new basis  with respective  set of  eigenvectors \{$\vec{P}$\} = [$\vec{P}_0, \vec{P}_1,..., \vec{P}_{424}$]   that span the database $\mathcal M$ is found.  

Consequently, any spectra  $\vec{S}_m$  can be expressed as a linear combination of the eigenvectors. Considering the change of variable previously described in  Sect. 2.1, the correspondent  expression  for the spectra $\vec{S}_m(X)$  in terms of the eigenvectors is : 

\begin{equation}
 \vec{S}_{m,i} ({\xi}_m, X) = \sum_{n=0}^{424}  \ \alpha_{n,i}^m \ \vec{P}_{n,i}(X) \ ; 
\  \  \   \  \  \    i=(I,Q,U,V),
\label{equ:pca2}
\end{equation}
where X expressed in km/s has replaced the wavelength value. Note  that  any Stokes spectra $\vec{S}_{m}$ can be already identified by an unique associated set of coefficients,   \{$\vec{\alpha}^m$\}=[$\alpha^m_0, \alpha^m_1,...,\alpha^m_{424}$].

\begin{figure*}[!htp]
\begin{center}
\vspace{0.25cm}


\resizebox{18cm}{!}{\includegraphics{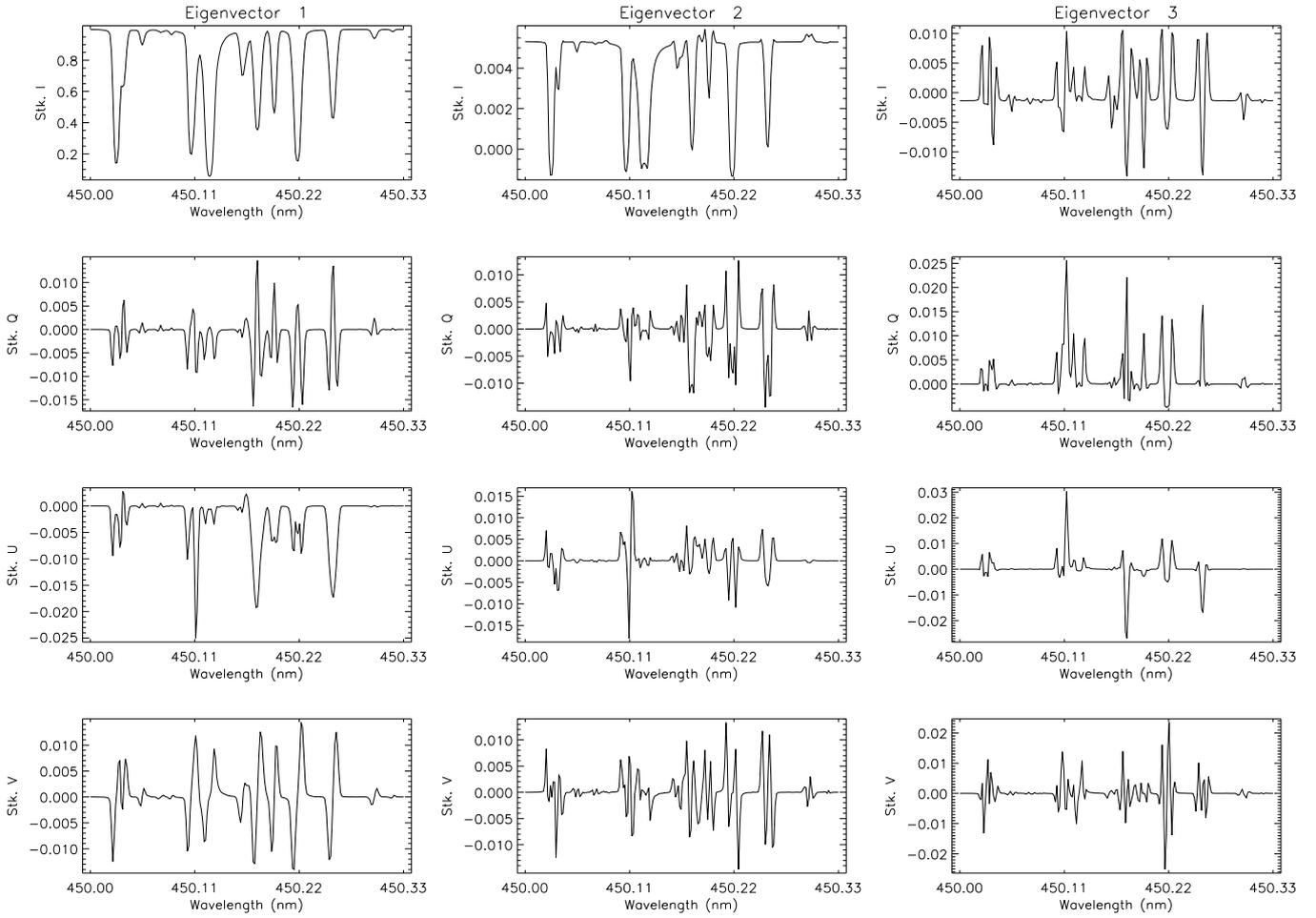}}
\vspace{0.25cm}
\caption{Plots of the first three eigenvectors, in the four Stokes parameters,  in a small spectral range of $\lambda$=[4500, 4503] \AA.} 
\label{fig:exa_ev}
\end{center}
\end{figure*}

In figure \ref{fig:exa_ev}, we show a small portion of the first three eigenvectors of the database constructed for T=4750 $^{\circ}$ K.

\begin{figure*}[!hbt]
\begin{center}
\vspace{0.25cm}
\resizebox{10cm}{!}{\includegraphics{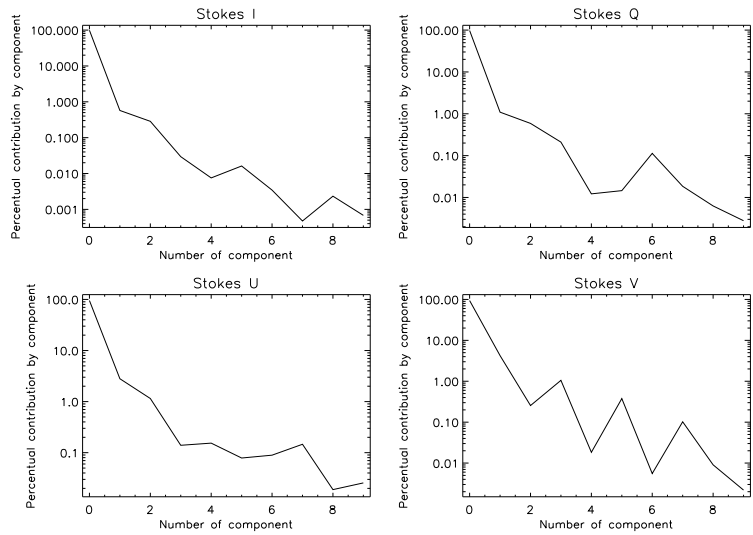}}
\vspace{0.25cm}
\caption{Percentage contribution of the first 10 components from Eq. (\ref{equ:pca2}) in logarithmic scale. Note that the contribution for the 10th component in the reconstruction of the original spectra is already at least inferior to 0.05 percent. }
\label{fig:comps}
\end{center}
\end{figure*}
\begin{figure*}[!hbt]
\begin{center}
\resizebox{16cm}{!}{\includegraphics{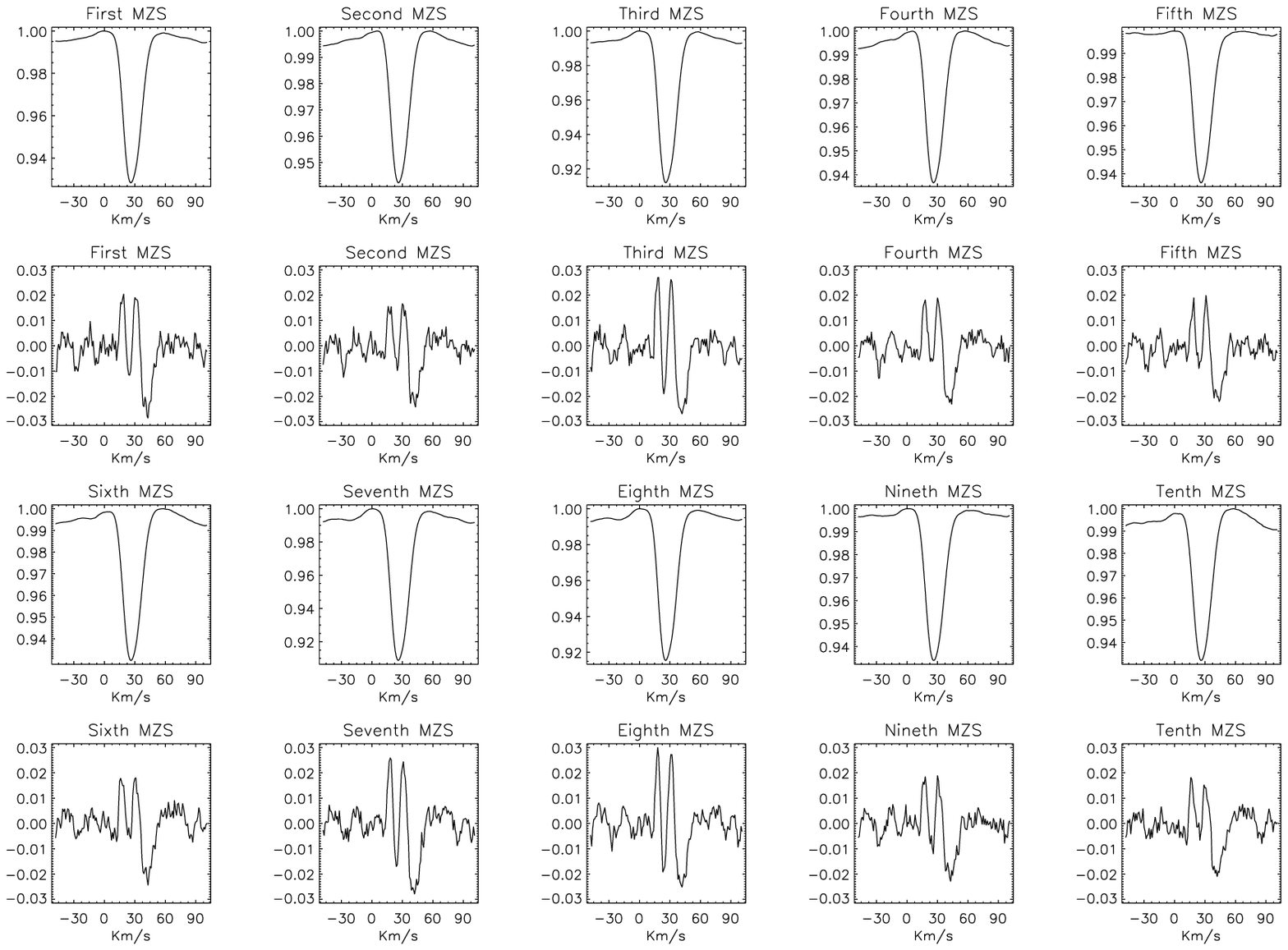}}

\hspace{5 cm}

\resizebox{16cm}{!}{\includegraphics{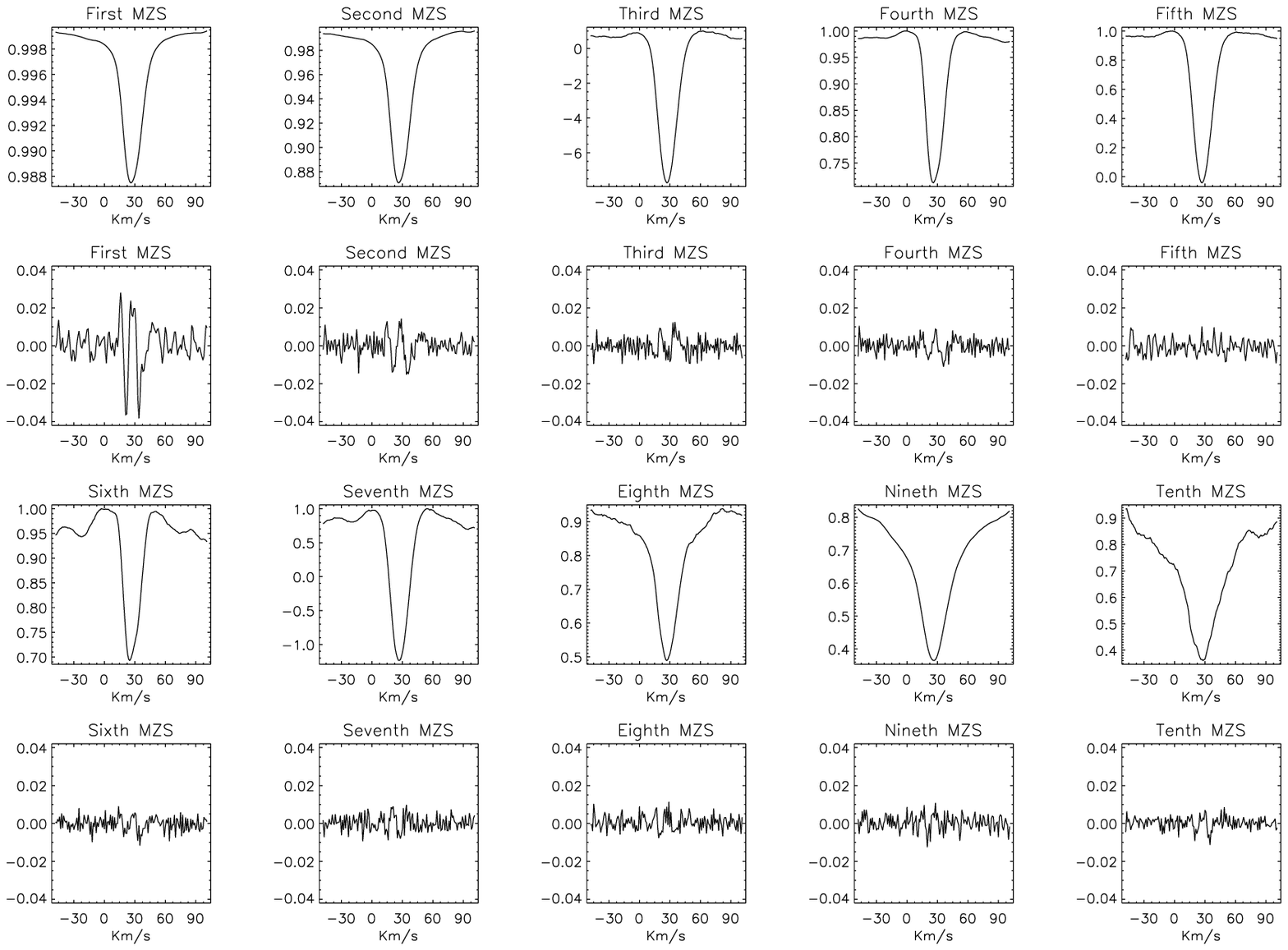}}
\caption{Comparison of the  MZS for the star IIpeg considering the absolute value of the eigenvectors (upper panels) and without it (lower panels).}
\label{fig:abs_iipeg}
\end{center}
\end{figure*}

\subsection{Space dimension reduction}
One of the reasons  why we employ  PCA  is because of its capability of data compression and space dimensions reduction. We mean that in Eq. (\ref{equ:pca2}) the equality  is reached only when $n_{max}$ = 424, but in practice $n_{max}$ could be dramatically reduced such that only the first components are considered. The consequent question is how to determine the number  of components really \emph{significants} i.e. how to determine $n_{max}$. To answer this question we have reconstructed all the profiles with 1 component, then with 2, and so on,  and we compare the reconstructed spectra with the original ones. The mean percentage contribution per component is plotted in Fig. \ref{fig:comps}. 

From Fig. \ref{fig:comps}, we can appreciate that rate of contribution of the components to the original spectra decreases faster than an exponential, permitting to truncate  the expansion in Eq. (\ref{equ:pca2}) to the first components. Given that the contribution for $n=9$  is already inferior to 0.001\% in the intensity Stokes parameter and inferior to 0.05\% in the polarized ones, we have decided to keep only the first ten terms in Eq. (\ref{equ:pca2}). The same number of components will be used when  obtaining the Multi-Zeeman-Signatures and when inverting them.   Moreover,  for any star with the same temperature that the one considered in the database, the same  ten eigenvectors  will be useful to retrieved the MZS.

At this point in the development of the technique  we have followed a typical approach with PCA. In the following we employ the eigenvectors as magnetic detectors such that combining them with the ZDI principles, the magnetic signature from the spectra will be extracted.

\subsection{The Multi-Zeeman-Signatures (MZS)}

Another of the advantages of the employment of PCA  apart from the data compression,  is that it permits to retrieve a set of coefficients \{$ \vec{\alpha}^m$ \} associated univocally to each one of the   spectra $\vec{S}_{m}$.  We will use then these coefficients to establish a relation to obtain the Multi-Zeeman-Signatures. 

Given that the eigenvectors have the property of being orthonormals, from Eq. (\ref{equ:pca2}) it is then trivial to obtain  the set of coefficients  for any of the synthetic Stokes spectra as:
 
\begin{equation}
\alpha_{n,i}^m= \vec{S}_{m,i}({\xi}_{m}, X) \cdot \ \vec{P}_{n,i}(X) ;     
\label{equ:alfas}
\end{equation}
where  n=[0,1,...,424] and  i=(I,Q,U,V). The relation obtained in Eq. (\ref{equ:alfas}) represents  the fundamental principle that will be employed to extract the magnetic signature from the observed spectra ($\vec{S}^{obs}$). In order to do it, the last step of the procedure is to consider the rotation velocity of the star and the associated Doppler effect, i.e. combine the MZS profiles with the ZDI principles.
 
It is important to mention that in order to equalize the spectral resolution of the observations with the one of the eigenvectors, we degraded the resolution of the eigenvectors to the spectral resolution of the observations.  This last  is done through a linear intrapolation similar the one described in Sect.  2.1. 

We proceed  now to  apply the same operation expressed in Eq. (\ref{equ:alfas}) to the observed spectra $\vec{S}^{obs}$, but when the eigenvectors  are Doppler shifted by an amount Y=$\Delta X$ : 

\begin{equation}
\alpha_{n,i}^{obs} (Y) = \vec{S}_{i}^{obs}(X) \ \cdot \ \vec{P}_{n,i}(X-Y).
\label{equ:alfas2}
\end{equation}

The vectorial product in the right side of in Eq. (\ref{equ:alfas2}) that permits to retrieve the $ \alpha  $-coefficients  at each considered $Y$-value is equivalent to a cross correlation function between the spectra and the magnetic detectors (the eigenvectors). 

On the other hand, it is convenient to mention that the expression found in Eq. (\ref{equ:alfas2}) can be replaced by another one where the eigenvectors \{$\vec{P} $\} are substituted by the absolute value of the same eigenvectors, denoted by \{ $\mid \vec{P} \mid $ \} : 

\begin{equation}
\alpha_{n,i}^{obs} (Y) = \vec{S}_{i}^{obs}(X) \  \cdot \ \mid \vec{P}_{n,i} \mid  (X-Y) .     
\label{equ:alfas_abs}
\end{equation}

Considering the whole range of Doppler velocities, Y=[$Y_0,Y_1,...,Y_{max}$], the general expression of the  Multi-Zeeman-Signature profiles ($\vec{P_{MZS}}$) is  given by: 

\begin{equation}
(\vec{P_{MZS}})_{n,i}(Y)= \left( \alpha_{n,i}^{obs} (Y_{0}),\ \alpha_{n,i}^{obs} (Y_{1}),\ ...\ ,\ \alpha_{n,i}^{obs} (Y_{max})  \right) .
\label{equ:Psp}
\end{equation}

\begin{figure*}[!ht]
\begin{center}
\resizebox{18cm}{!}{\includegraphics{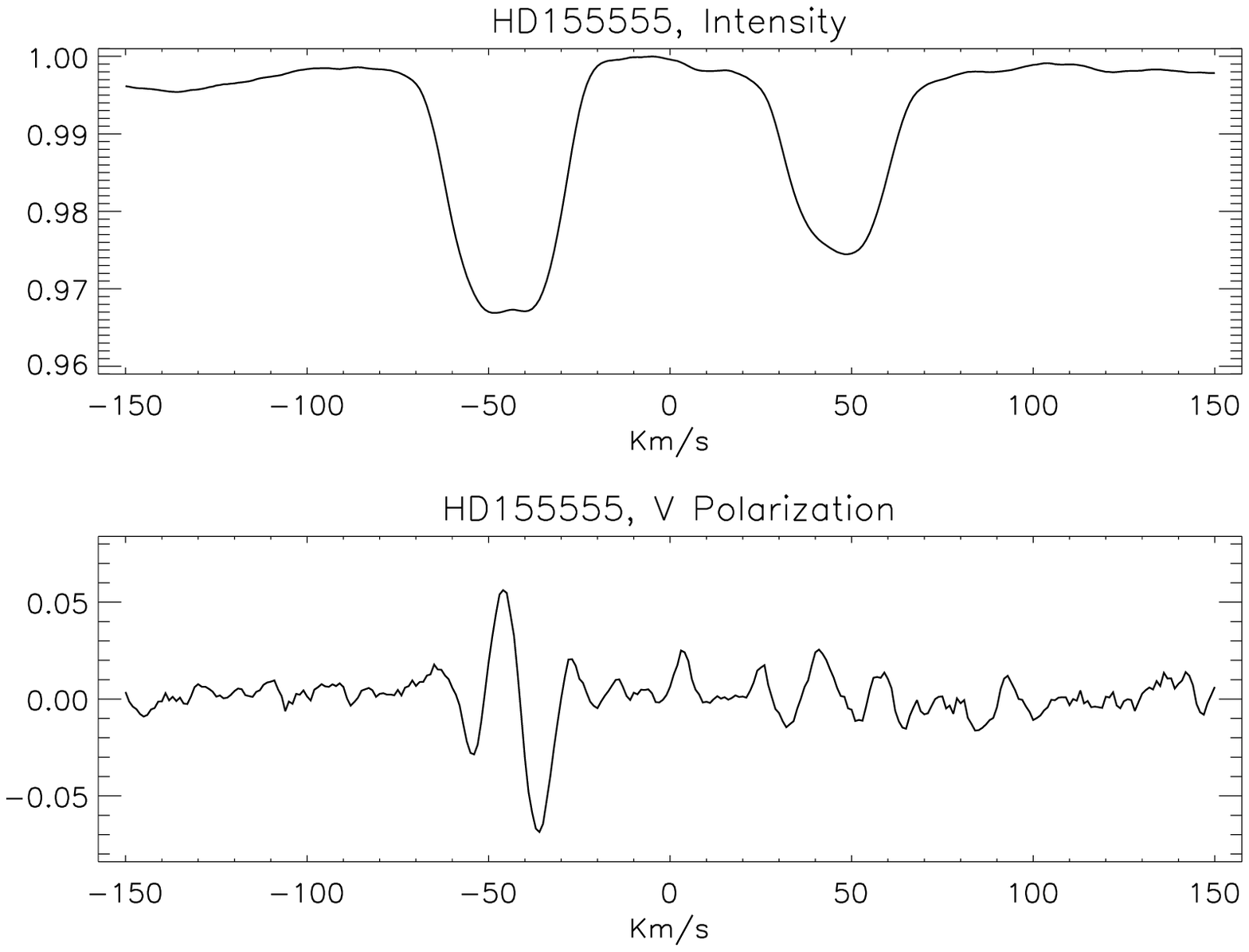}{\includegraphics{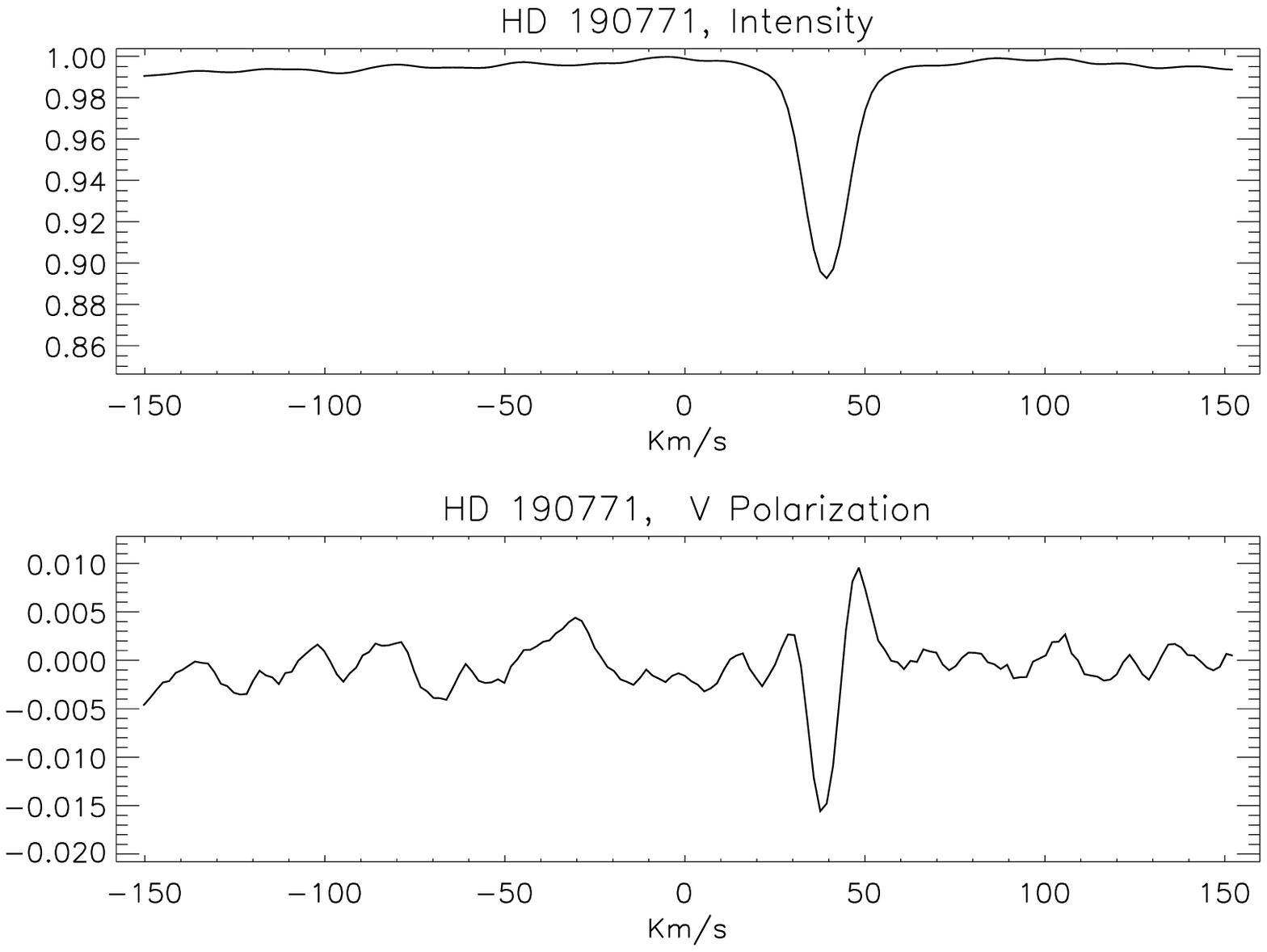} }}
\caption{Examples of normalized intensity and circular  polarization profiles of two solar type stars obtained with the presented PCA-ZDI technique. The left plots shows the results for the {\it HD 155555 binary system}  observed at the the AAT telescope (Australia) in April 2007. The right plots correspond to the results for the {\it HD 190771} star observed at the Bernard Lyot telescope in August 2007.}
\label{fig:detec}
\end{center}
\end{figure*}

Note that for each Stokes parameter, (I,Q,U,V),  425 Multi-Zeeman-Signatures are retrieved. Since the eigenvectors are orthonormals, then each one of the 425 Multi-Zeeman-Signatures are in principle independents, making possible to perform independent  analysis with each one of them by separate.


\subsection{Magnetic detections}

Once presented the technique, we will now apply it to the real observed spectra. We will employ  the eigenvectors obtained for the case of  single points in the stellar surface,   however this  fact does not limit the detection of the global magnetic field. 

 We will first compare the MZS obtained from Eq. 10 to those retrieved following  Eq. 11. 
Let  $v_{max}$ and  $v_{min}$  denote the  maximum and  minimum projected rotation velocities of the star $vsini$, such that  the star velocities span  as  $V^{rot}=[v_{min},...,v_{max}]$.  If the considered Doppler velocity Y does not coincide with any of the $V^{rot}$  values, the  intensity MZS profile recovers the continuum value, and the polarized MZS profile an aleatory sequence of  values around zero. 

Contrary, if Y coincides with one of  the $V^{rot}$  values, the magnetic signature in the polarized parameters and the total intensity profile are retrieved. In Fig. \ref{fig:abs_iipeg} we compare  the first nine  MZS  considering the absolute value of the eigenvectors (upper panels) to the  first nine  MZS when the absolute value is not included (lower panels). The data correspond to the \emph{IIpeg} star observed in August 2007 with the Bernard Lyot telescope, at the Pic du Midi Observatory, using the  NARVAL polarimeter.

From  Fig. \ref{fig:abs_iipeg}, we appreciate that when the absolute value is included the nine polarized MSZ are clearly detected with a high S/N ratio. If  the absolute value is not considered the level of the first polarized MZS is clearly superior to the noise level but the level of the MZS began to decrease for the rest of eigenvectors.

A discussion about how to take advantage of this multiple Multi-Zeeman-Signature detections will be presented in a forthcoming paper.  In the following, we prefer to add the nine individual Multi-Zeeman-Signatures in order to obtain a total profile per Stokes parameter.

In the last subsection we have showed that the expansion in Eq. (\ref{equ:pca2}) can be reduced to the first ten components. We then  add  the contribution from these  components to find the final expression that will be employed to retrieve the Multi-Zeeman-Signature: 

\begin{equation}
(\vec{P_{MZS}})_{i}(Y)= \sum_{n=0}^{9} \left( \alpha_{n,i}^{obs} (Y_{0}),\ \alpha_{n,i}^{obs} (Y_{1}),\ ...\ ,\ \alpha_{n,i}^{obs} (Y_{max})  \right). 
\label{equ:Psp}
\end{equation}

Considering the absolute value of the eigenvectors and following the procedure of the last equation, in Fig. \ref{fig:detec}, we show two examples of the MZS profiles  obtained in intensity and in  circular polarization of two cool stars.  The binary system HD 155555, left panels, was observed in April 2007 with the Anglo Australian Telescope, at the AAO observatory, using the SEMPOL polarimeter. The solar type star HD 190771, right panels,  was observed in August 2007 with the Bernard Lyot telescope, at the Pic du Midi Observatory, using the  NARVAL polarimeter. 

The detection of magnetic fields in this three stars have been also found through  the LSD method. In fact,  the MZS profiles from Fig. \ref{fig:detec} seems  similars in shape to those retrieved with the LSD method (Dunstone et al. 2008, Petit et al. 2008). Nevertheless a proper comparison has not been done to distinguish possible differences in the profiles and how important could be those differences when retrieving the magnetic field.

In any case, for the moment, the PCA-ZDI technique has been applied with success to retrieve mostly the circular states of polarization in cool stars (Ram\' irez V\'elez et al. 2006). In fact this is in part  due to the absence of representative samples of linear data to analyse, so  we leave  for a future work  studies where the linear states of polarization can be incorporated in the detection and measurement of stellar magnetic fields. 

\subsection{Inversions of the Multi-Zeeman-Signatures}

In this section we intend to show that it is possible to infer the magnetic field  through the inversion of the MZS profiles. We will place a simplistic scenario where again we consider local Stokes profiles from different surface elements. We further assume that there is a punctual magnetic field in the surface elements. From now on, the MZS to which we refer will be obtained from the local Stokes profiles through Eq. (13).  Given a MZS, the inversion exercise  consist in find  the position of the surface element and the magnetic field vector.

The local Stokes profiles are calculated for a small spectral  range [5400,5500] \AA.  The parameters of the Stokes profiles, and thus those of the MZS,  varies randomly in the following ranges : the magnetic field strength [0,10] kG, the inclination field [0,90]$^{\circ}$, the magnetic azimuthal angle [0,180]$^{\circ}$ and the projected position  of the surface element ($\mu$) [0,1].

Since the inversions are performed in the space of the MZS we have produced 6000 MZS that serves as database where we tested the inversion of a set of 600 MZS.
	
As mentioned previously there are two alternatives to produce  MZS, namely, whether or not is considered the the absolute value of the eigenvectors (Eqs. 10 and 11). We will test both alternatives and we will show that whether or not is included the absolute value, the inversions are correct. 

\begin{figure}[!hb]
\begin{center}
\resizebox{9cm}{!}{\includegraphics{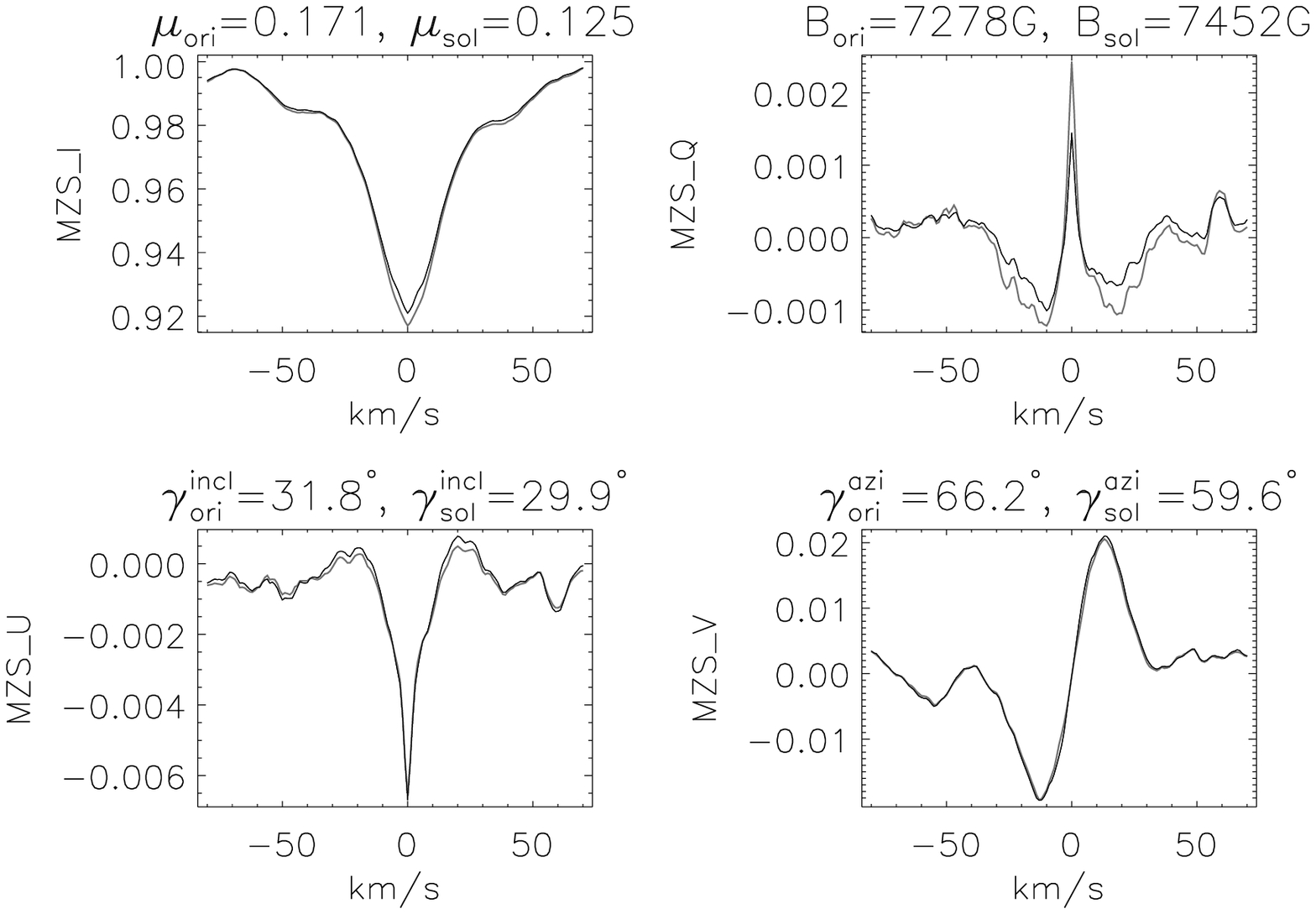}}
\resizebox{9cm}{!}{\includegraphics{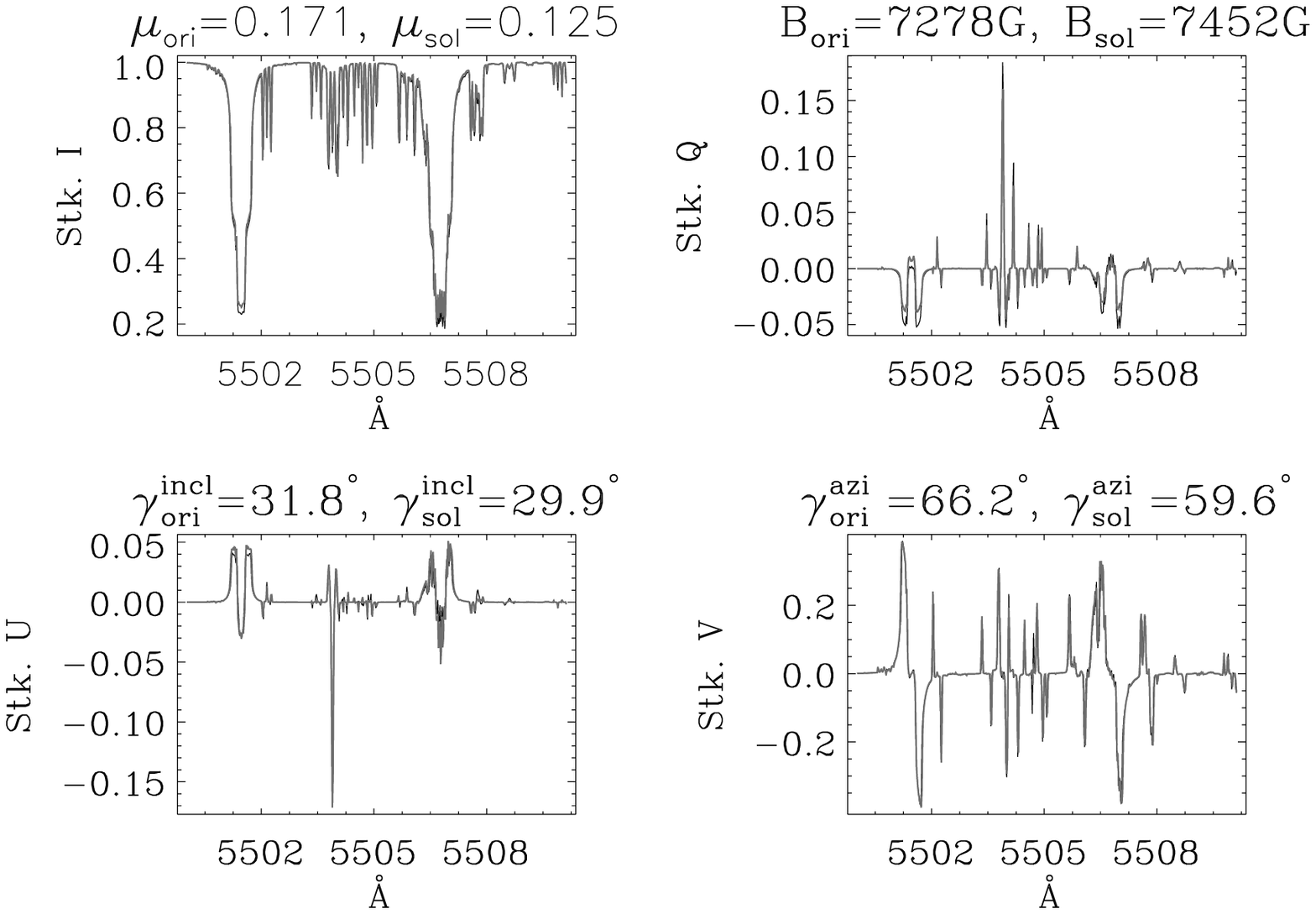}}
\caption{In black the original profiles, in gray the retrieved solution. Upper panels, examples of the MZS when the absolute value of the eigenvectors is considered. Lower panels, a small spectral range of 10 \AA \ of the Stokes profiles associated to the MZSs. } 
\label{fig:examp}
\end{center}
\end{figure}

We first present the case where the absolute value of the eigenvectors is employed to retrieve the MZS. In upper panel of Fig. \ref{fig:examp} we show  an example of a synthetic MZS (in black color) and the respective MZS solution (in gray color).  The inversion is correct not only because the similarity between the MZSs but because the values of the parameters solution are close the originals ones (see the title in each panel).  Moreover, given that any MZS is associated to a stellar spectrum, in the lower panels of Fig. \ref{fig:examp} we  corroborate that, as expected,  the Stokes profiles associated to the MZS solution fit enough well  the originals ones.  

\begin{figure}[!hb]
\begin{center}
\resizebox{9cm}{!}{\includegraphics{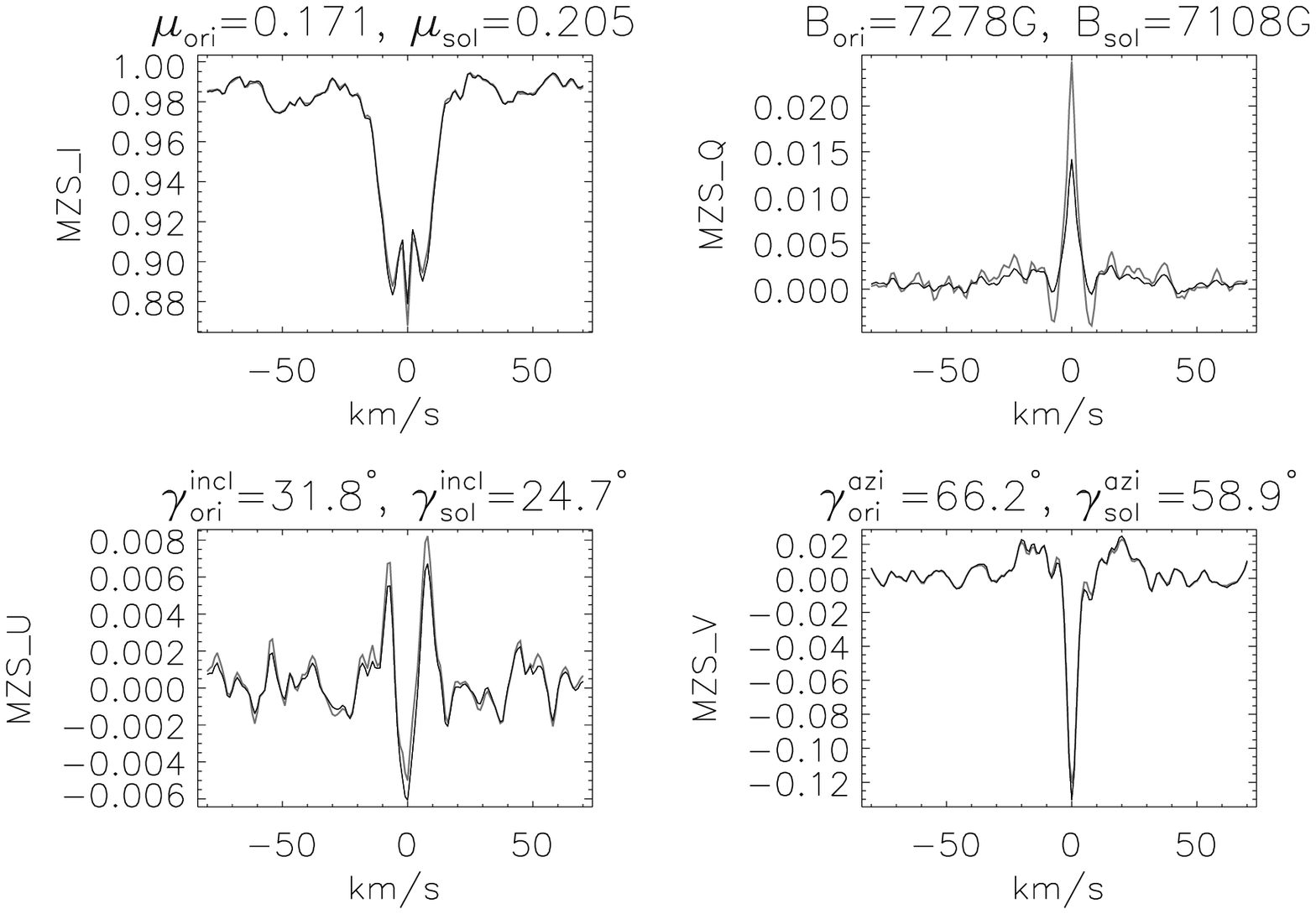}}
\resizebox{9cm}{!}{\includegraphics{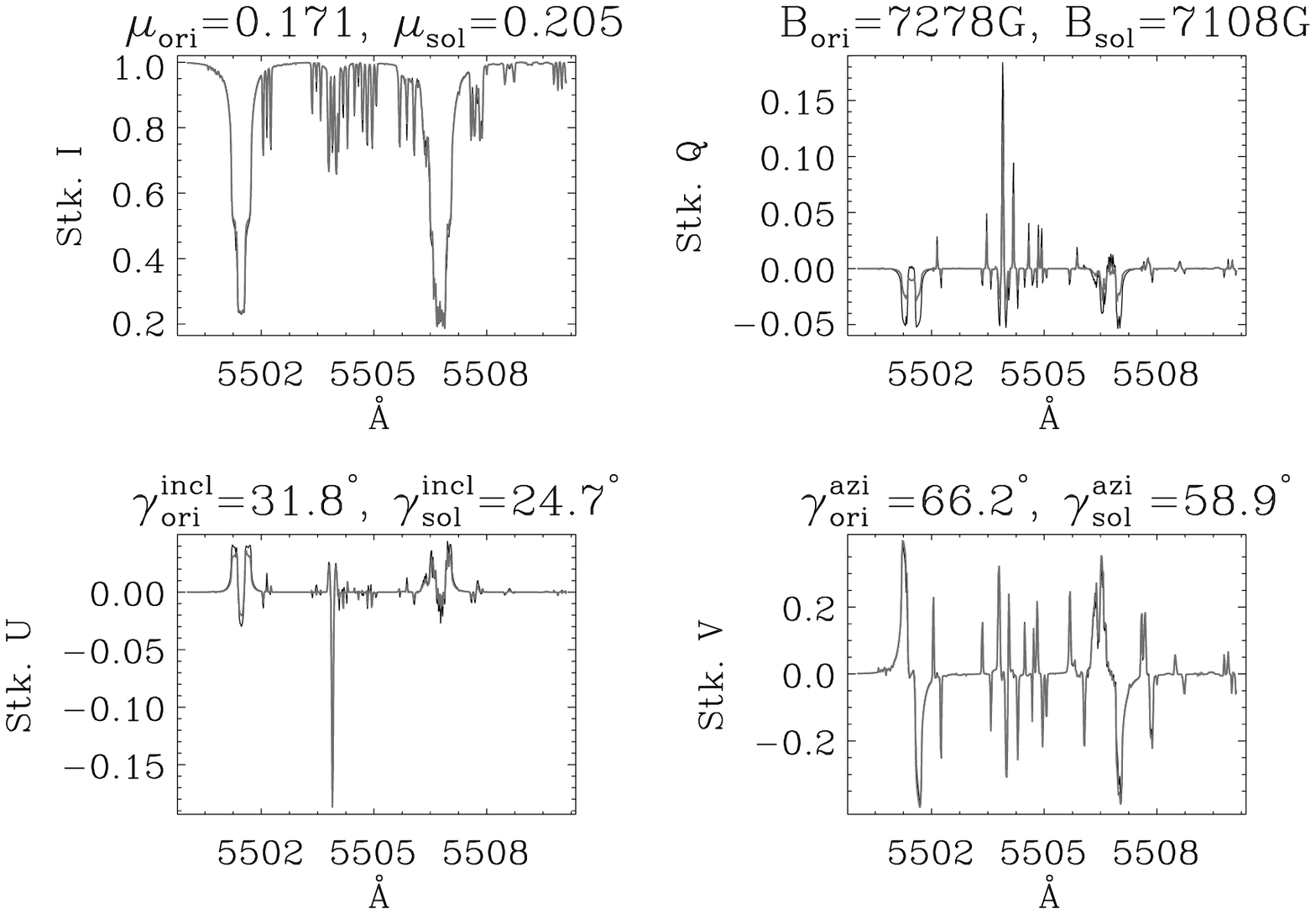}}
\caption{Same as in Fig. \ref{fig:examp} but when the absolute values of the eigenvectors is not considered to produce the MZS.}
\label{fig:examp2}
\end{center}
\end{figure}

In Fig. \ref{fig:examp2} we show the MZS associated to the same combination of parameters as before but for the case where the absolute value is not considered. The solution parameters in this case  are also close to the original ones (upper panels) and the associated solution Stokes profiles gives a good fit to the original Stokes profiles (lower panels). 

\begin{figure}[!ht]
\begin{center}
\resizebox{9cm}{!}{\includegraphics{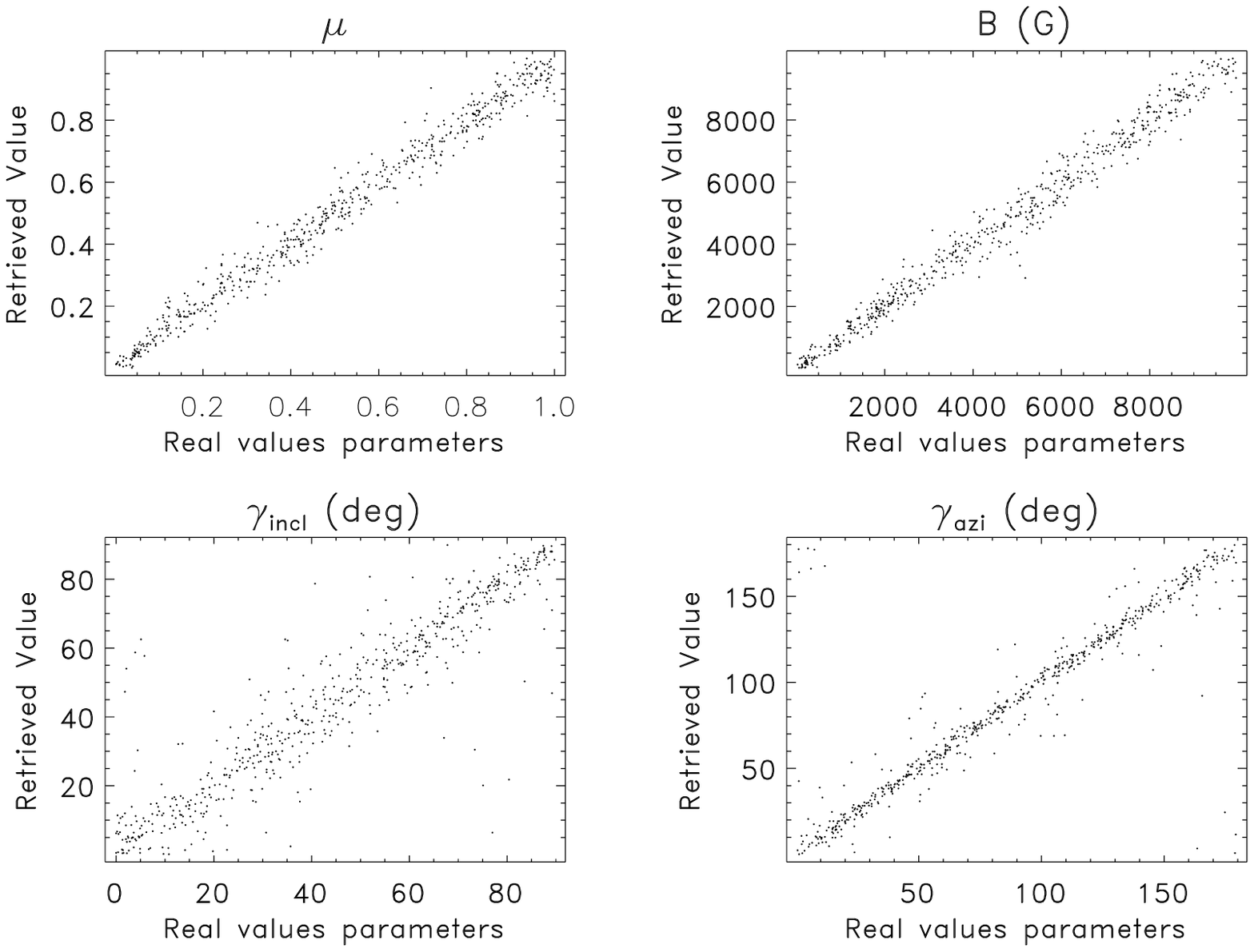}}
\resizebox{9cm}{!}{\includegraphics{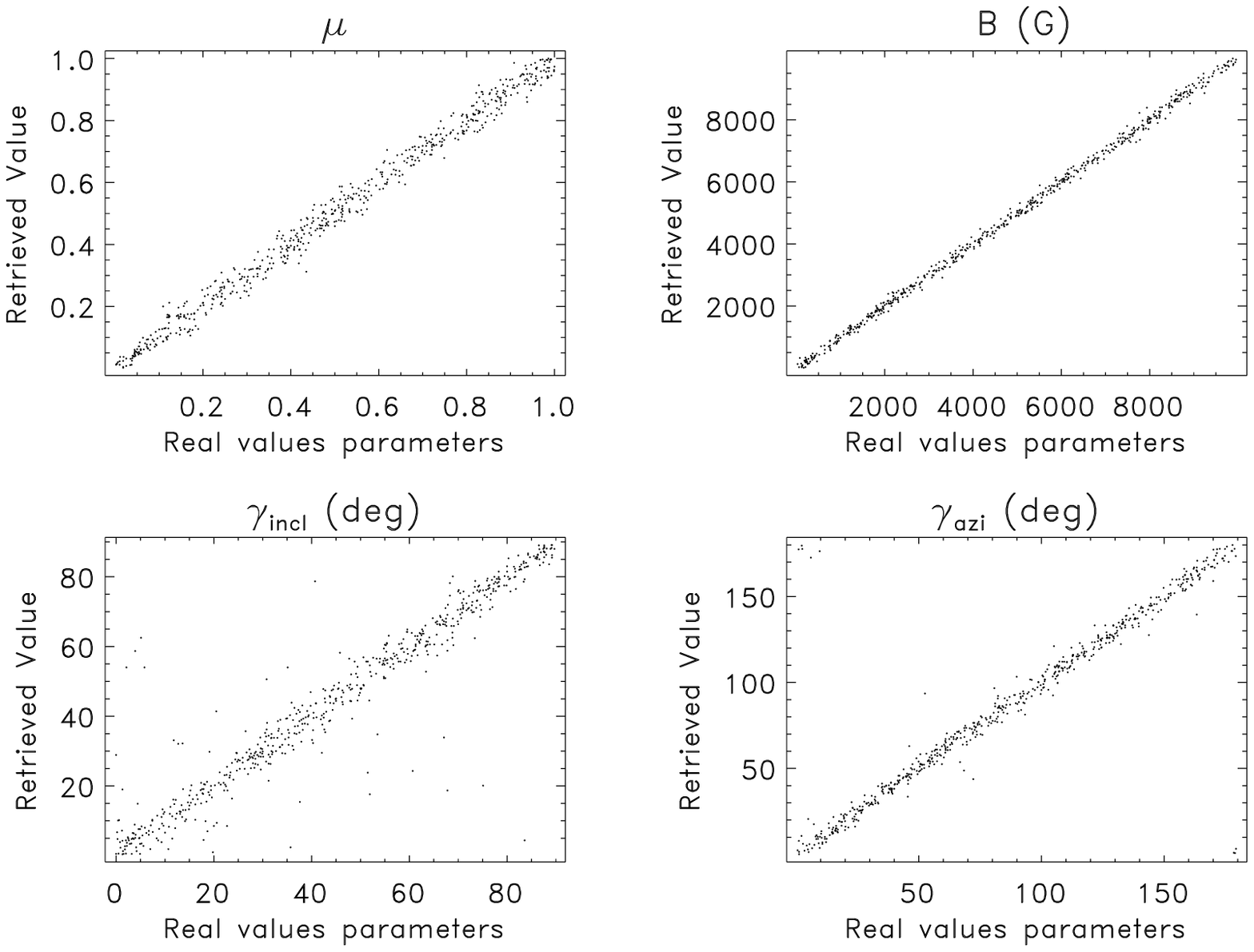}}
\caption{
Inversions results of the entire set of MZS profiles. The upper (lower) panels are the results for the case when the absolute value of the eigenvectors is (not) included to produce the MZS. In both cases the parameters are properly retrieved.}
\label{fig:inv_psp}
\end{center}
\end{figure}

Finally, we present the results of the inversions of the 600 MZS. In Fig. \ref{fig:inv_psp}, the upper panels correspond the case when the absolute value of the eigenvectors is included to produce the MZS and the lower panels correspond to the case when is not included. 

These results  show that in both cases,  the magnetic field vector is correctly retrieved despite the position in the stellar surface, for all  strength fields and  orientations.  

The final conclusion in this section is that the Multi-Zeeman-Signature  represents not only an effective tool for the detection of stellar magnetic fields, but also that inversions in the space of the Multi-Zeeman-Signatures can be directly performed.

\section {Conclusions}

In this article, we have focused in  the multi line analysis of spectropolarimetric data, with emphasis in the detection and inference of stellar magnetic fields.   The use of multiple lines is required to increase the signal to noise ratio in those cases where the polarized signals in individual spectral lines are significantly inferiors to the noise level, remaining hidden for  analysis purposes.  This is typically the case, for instance, in observations of cool Stars.

To deal with this observational constraint, we have presented the development of  two new  approaches that serve both of them to extract a polarized signal from a mixture of  several individual spectral lines.  While employing any of the proposed multi-line techniques, namely the pseudo line or the PCA-ZDI approach,  we remark three main advantages: (1) the contributions of all the spectral lines to  the final polarization signal are included  -despite the similarity or not in the individual shape profiles-,  (2) it is   a valid method  for the detection of the circular and the linear states of polarization and (3) it is not limited to weak or to strong magnetic field regimes.

In the case of the pseudo line, we have presented an inversion code that builds a database of pseudo lines each one attached to a particular atmospheric model.

Initially, considering an ideal scenario where the magnetic field is the only free variable of the atmospheric model,   we have found that the results obtained from the inversions of the pseudo line are as goods as those obtained from individual spectral lines.  We have also shown that it is expected that in the case of real observed spectra (with a given noise level), the best results in any of the magnetic components are retrieved with the inversions of the pseudo line.   Finally, we have showed that the atmospheric model, in addition to the magnetic field,  can also be inferred  through  the inversions of the pseudo line.

\vspace{0.25cm}
In the case of the Multi-Zeeman-Signatures, we discussed in detail the development of the technique, and we applied it to real observed data in order to illustrate  the type of profile obtained with the PCA-ZDI technique.  

The circular polarized MZS profiles, like those in Fig. \ref{fig:detec} where the polarized signal level is clearly superior to the noise level, represent in a  unambiguously way a detection of the magnetic field present in the stellar object.  

By construction, the Multi-Zeeman-Signatures  contain the information of the magnetic field  and atmospheric model. To  verify this,  we have performed inversions in a synthetic set of  MZS profiles.  The results obtained after the inversions  show that the all the considered parameters in the atmospheric model are correctly retrieved. 

It is pertinent to mention that some refinements have to be done to the technique in order to achieve magnetic fields measurements through the inversion of the MZS profiles. In particular, the Doppler broadening effect due to the rotation of the stars and the case of inversions of continuum fields distributions over the stellar surface will  be presented in forthcoming papers. For the moment,  we have settled the basis of the approach founding that PCA-ZDI is a robust technique for the analysis of stellar magnetic magnetic fields.

The final conclusion of this work is that when the inversion of the magnetic polarized signal are done through  the same method used to construct them, the physical parameters that determine the line formation process can be properly recovered, included in particular, but only, the stellar magnetic field.



\bibliographystyle{/home/julio/articulos/bibtex/aa}

\end{document}